\documentclass[sigconf]{acmart}

\AtBeginDocument{%
  \providecommand\BibTeX{{%
    \normalfont B\kern-0.5em{\scshape i\kern-0.25em b}\kern-0.8em\TeX}}}

\copyrightyear{2024} 
\acmYear{2024} 
\setcopyright{licensedothergov}\acmConference[IDE '24]{2024 First IDE Workshop}{April 20, 2024}{Lisbon, Portugal}
\acmBooktitle{2024 First IDE Workshop (IDE '24), April 20, 2024, Lisbon, Portugal}
\acmDOI{10.1145/3643796.3648444}
\acmISBN{979-8-4007-0580-9/24/04}

\acmConference[ICSE 2024]{46th International Conference on Software Engineering}{April 2024}{Lisbon, Portugal}

\begin{document}

\title{Challenges of Processing Data Clumps within Plugin Architectures of Integrated Development Environment}

\author{Nils Baumgartner}
\email{nils.baumgartner@uni-osnabrueck.de}
\orcid{0000-0002-0474-8214}
\affiliation{%
  \institution{Software Engineering Research Group, School of Mathematics/Computer Science/Physics, Osnabrück University}
  \country{Germany}
  \postcode{49074}
}

\author{Elke Pulvermüller}
\email{elke.pulvermueller@uni-osnabrueck.de}
\orcid{0009-0000-8225-7261}
\affiliation{%
  \institution{Software Engineering Research Group, School of Mathematics/Computer Science/Physics, Osnabrück University}
  \country{Germany}
  \postcode{49074}
}



\begin{abstract}
  In this study, we explore advanced strategies for enhancing software quality by detecting and refactoring data clumps, special types of code smells. Our approach transcends the capabilities of integrated development environments, utilizing a novel method that separates the detection of data clumps from the source access. This method facilitates data clump processing. We introduce a command-line interface plugin to support this novel method of processing data clumps. This research highlights the efficacy of modularized algorithms and advocates their integration into continuous workflows, promising enhanced code quality and efficient project management across various programming and integrated development environments.
\end{abstract}



\received{9 November 2023}
\received[revised]{16 January 2024}


  
\maketitle

\section{Introduction}

Software errors, as highlighted by Brown et al. \cite{CostOfMaintenance}, significantly increase development costs. "Code smells," defined by Fowler et al. \cite{Fowler1999}, signify poorly structured but functional code. As Fowler explains, these code smells can be detected and refactored. Code smells are diverse and of varying complexity, and their identification and correction vary significantly. Several of these elements are specific to a method or line of code and only require the inspection of a single file. However, others, such as data clumps, are more complex and constitute significant portions of a project.

Due to different programming languages and integrated development environment (IDE) tools, new plugins need to be created, developed, and maintained for multiple IDEs and programming languages based on current interfaces. Detecting and refactoring data clumps across diverse environments poses two primary challenges:
\begin{itemize}
    \item Upholding a synchronized algorithm for precise detection across various environments such as IDEs, continuous integration and continuous delivery (CI/CD) pipelines, large-scale automated projects, and web applications.
    \item Ensuring consistent support for refactoring techniques across these diverse environments.
\end{itemize}

\section{Background}

A data clump is a group of variables typically found in software projects, suggesting overlooked, deeper structures. Their varying sequences make their detection challenging, necessitating an analysis of the entire project. They can occur within a file and across multiple files, making them difficult to identify and resolve. The definition of data clumps has been improved by Zhang et al. in \cite{ImprovedCodeSmellDefinition}, helping researchers take an automated approach to refactoring and improving code quality and maintainability.

\begin{figure*}[htbp]
    \centering
    \includegraphics[width=\textwidth]{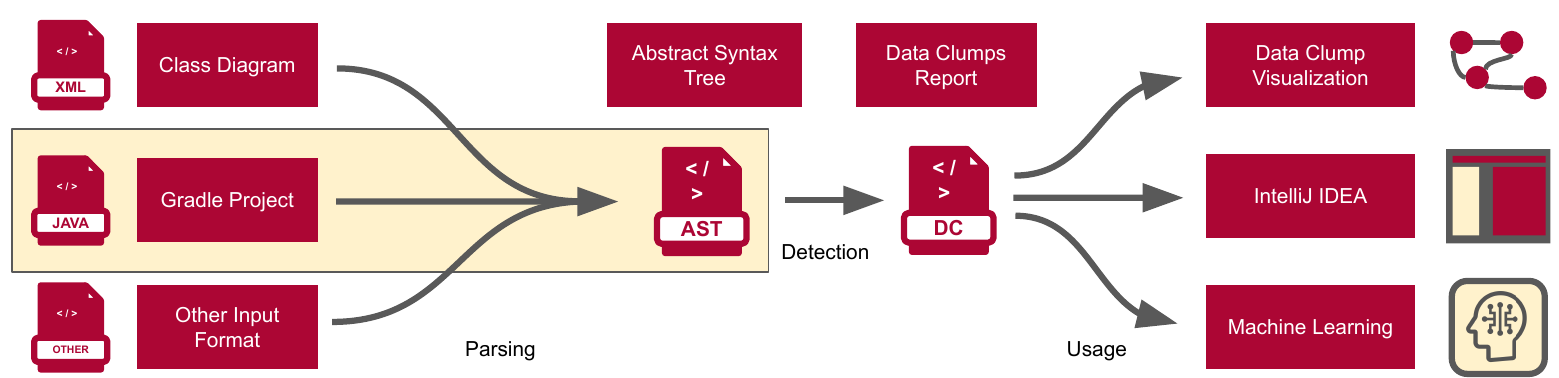}
    \caption{Schematic Diagram of the Architecture for Artifacts Processing and Data Clump Detection}
    \Description{Schematic Diagram of the Architecture for Artifacts Processing and Data Clump Detection}
    \label{fig:splitApproach}
\end{figure*}

In terms of software quality, the use of automation strategies has increased significantly in recent years \cite{improveSoftwareQualityThroughAutomation}. Various methods can support developers in their endeavors, such as IDEs, which are not only programming platforms but also powerful development support tools with features such as automatic refactoring \cite{automateRefactorings}. A notable advantage of an IDE is its ability to preload crucial libraries and additional dependencies, helping streamline developers' workflows, allowing them to focus on more complicated aspects of coding.

In contrast, designing standalone tools allows developers to explore novel approaches and develop innovative methods. These autonomous tools are designed to be compatible with appropriate computer systems, adding another layer of versatility to the developer’s toolkit. Nevertheless, these tools face a significant challenge regarding the programmatic parsing of the source code. This obstacle can be mitigated using sophisticated build tools, such as Gradle, Ant, and Maven, which are conduits for seamlessly downloading and installing the prerequisites for various software projects. While standalone tools have their challenges, they highlight the need for efficient code analysis methods and facilitate static code analysis.

A key strategy for improving software quality is to statistically check the source code. This process can be expedited by converting the source files into more navigable structures, such as an abstract syntax tree (AST) or a concrete syntax tree (CST), simplifying the nature of the raw code \cite{refactoringBot}. An alternative but equally effective approach is to analyze compiled files. However, this method presents numerous challenges, primarily because these compiled artifacts are not consistently uploaded to version control repositories, sometimes limiting their accessibility for quality improvement.

\section{Related Work}

When reviewing relevant work, it is crucial to differentiate between our previous contributions and current advancements. Previously, our focus was on integrating code smell detection, specifically data clumps, within the IntelliJ IDE, leveraging its inherent capabilities. This section compares our past work with existing tools and methodologies to explain the evolution and significance of our current research.

Our previous research concentrated on an IntelliJ plugin, LCSD, designed for the real-time detection and semi-automatic refactoring of data clumps. The plugin capitalized on IntelliJ’s built-in interfaces, offering a seamless user experience within this specific IDE. Its effectiveness was rooted in the direct and immediate feedback mechanism, which was a key feature for developers working within the IntelliJ environment \cite{baumgartner2023live}.

In contrast with other tools, such as PMD \cite{PMD}, our IntelliJ plugin offered unique advantages in real-time detection and the ability to select data clumps and refactor them automatically by providing a name for the newly created class. However, it also had limitations, primarily its confinement to the IntelliJ IDE, which restricted its applicability in diverse development scenarios.

In our previous research, we presented an IntelliJ plugin that could detect data clumps in real-time and provide a semi-automatic refactoring process \cite{baumgartner2023live}. This plugin uses IntelliJ’s built-in interfaces to parse and refactor a given project.

PMD is a static analysis tool for source code supported by IntelliJ, which can automate improvements \cite{PMD}. However, its current limitations include the need to define enhancements or error detections for each file individually. As a result, PMD can't autonomously resolve data clumps, which require a project-wide comparison. It also can't detect fully qualified class names or external library dependencies, as it only analyzes the given code base.

Reflective Refactoring ($R^2$) \cite{scriptingParametricRefactorings} is a Java package for automating the creation of classic design patterns. The tool can detect and refactor for 18 of the 23 Gang-of-Four design patterns.

Qodana, developed by JetBrains, is a comprehensive static code analysis tool tailored for CI/CD pipelines.\footnote{https://www.jetbrains.com/qodana/} Qodana offers a holistic view of code quality and uniquely analyzes dependencies. It excels in providing refactoring suggestions during the CI/CD process, bridging the gap between code analysis and actionable improvement.

\section{Approach}

Previously, we discussed the benefits of an IDE and developed a plugin that could detect data clumps in real-time and generate initial refactoring suggestions \cite{baumgartner2023live}. However, this process potentially involved further challenges, such as interacting with users so that they could learn from these design errors or selecting only desired data clumps. Therefore, an automated approach could help minimize human errors \cite{influenceOfHumanFactorsInSoftwareQuality}.

Our approach separates detecting data clumps from accessing the source data to automate improvement suggestions using data clump refactoring. An advantage of this approach is the potential to create an autonomous bot, similar to work by Wyrich and Bogner \cite{refactoringBot}, to help to create high-level refactoring, as described in \cite{howWeRefactor}.

A schematic representation of our approach is depicted in Figure \ref{fig:splitApproach}. The figure illustrates a workflow for detecting data clumps in software projects. Different input formats, such as XML, Java, and other file types, undergo a parsing process to generate an AST. The AST aids in data clump detection, resulting in a data clumps report. The findings can be visualized, used in the IntelliJ IDEA environment, or further analyzed using  machine learning techniques.

For each new programming language or IDE, we only need to
develop a plugin that reads the source files and parses them into a suitable format or uses a language server protocol. In our approach, we developed a command-line interface (CLI) plugin for IntelliJ based on the IntelliJ framework \cite{intellijpluginframework}. The CLI plugin can then be called via a Docker supporting the programmatic detection and refactoring of data clumps. The CLI tool uses two arguments: input and output. Input specifies the path to the project, and output specifies the destination folder for the parsed files. We used the output for our standalone detection tool.

Our novel solution pivots away from IDE-specific and exclusive plugins toward a more centralized approach. The idea is to encapsulate detection logic in a shared resource, for example in an npm package, which various IDE plugins can then utilize for an IDE plugin. Currently, our solution uses the CLI plugin for IntelliJ, as it enables our IDE plugin to be reused. We are working on splitting the current approach to centralize detection solely using an AST, as the CLI plugin only parses the project source code into the AST needed for the detection.

Our approach provides flexibility, as our tool can be extended to various IDEs and environments without the need to duplicate the core logic. The second is centralized maintenance: updates or improvements to the detection algorithm only need to be performed in one place, simplifying maintenance and ensuring consistency across platforms. However, this approach does have disadvantages. One is integration complexity: merging different run-time environments and ensuring compatibility across various IDEs can be challenging. Another is potential latency in feedback. Unlike the direct, real-time feedback in the IntelliJ plugin, this approach could introduce a delay in the detection and refactoring feedback loop, depending on the implementation. Additionally, we developed a project parser using PMD. Although PMD provided the files in an accessible format, it did not consistently and reliably identify dependencies and qualified names of classes and packages. This additional parser is suitable for a broad analysis of projects in which build tools such as Gradle, Ant, or Maven are not used, and only the source code is available.

\bigskip

To detect data clumps, we initiated the CLI plugin with specific arguments and used IntelliJ’s \textit{Repository Opener} to open the project. IntelliJ then employs Gradle for dependencies and ensures the project is fully loaded using IntelliJ’s \textit{ApplicationManager}. We extracted data from each Java module for clump detection within that module, simplifying refactoring. However, this method could miss inter-module clumps. Therefore, for each module, we parsed the AST of the source files and related library details. The AST information from libraries is marked as \textit{auxclass} to differentiate and focus on source files during the detection phase.

\section{Challenges and Possible Solutions}

Our previous approach relied on a centralized codebase to support data clump detection and refactoring in various IDEs, such as WebStorm (for JavaScript) or CLion (for C). We utilized a monorepo in GitHub, coupled with a build tool that automatically constructed plugins for each IDE and published them. This method ensured the synchronization of our main logic across different environments. Although effective in certain scenarios, this approach had several limitations. We faced significant challenges concerning IDEs that did not support plugins developed in Java. A potential solution was to shift to a cloud-based system. However, this approach raised concerns regarding data privacy and the complexity of managing cross-platform compatibility.

In our updated strategy, we focused on segregating the detection logic into a separate package, preferably an npm package, to facilitate integration with web applications. This approach still requires a plugin to be developed for each IDE. However, these plugins now import the required detection package. The plugin’s role is to process the project’s source code, convert it into an AST, detect data clumps and generate refactoring suggestions using a unified method list. Although this state has not been fully achieved yet, it presents significant advantages. It allows us to offer support at multiple interaction points, including web interfaces, IDEs, and CI/CD pipelines. This versatility enhances user accessibility and experience across different platforms and development contexts.

\bigskip

When we developed the previous plugin for the live detection of data clumps, we realized that several data clumps extended over larger areas and were connected to numerous files. The difficulty with our previous plugin was that the user had too little control over whether refactoring should occur. In the plugin, all the related data clumps were correctly refactored, so the user could not choose whether to exclude any of them. We developed a tool\footnote{https://github.com/FireboltCasters/data-clumps-visualizer} for visualizing data clumps to bring users closer to the dimensions of the data clumps. This tool can be accessed in the browser and does not require a backend; the user merely needs to insert the detected data clumps report. This process allows users to view the report independently of an IDE. The data clusters are represented in a graph, where the tool presents the files, classes, methods, and parameters as nodes. These nodes are connected, and data clumps exist between them. The package can be used programmatically, facilitating integration into other environments and supporting Node Package Manager (NPM)\footnote{https://www.npmjs.com/} packages. We identify the challenges of integrating visualizations into the IDE as a plugin while retaining the advantage of centralized maintenance.

~\\
~\\
\noindent \textbf{Challenges:} Several challenges arise when developing IDE plugins and standalone tools for detecting and refactoring data clumps:

\begin{enumerate}
    \item Compatibility with different IDEs and environments
    \item Performance and efficiency in large code bases or complex structures
    \item Creating an intuitive and interactive user interface for visualizing data clumps or other code smells or design smells
    \item Adapting the tool to different programming languages with unique syntaxes and structures
    \item Integrating the tool into existing development workflows and CI/CD pipelines
    \item Ensuring scalability and maintainability with project growth
    \item Maintaining data privacy and security, particularly in cloud-based or external processing scenarios
    \item Keeping the tool relevant and useful as programming languages and development practices evolve
\end{enumerate}

\noindent \textbf{Proposed Solutions:} The following solutions are proposed for each of the challenges identified:

\begin{enumerate}
    \item Develop a core library or package for consistent application across different IDEs
    \item Optimize algorithms for data clump detection and refactoring and consider asynchronous processing
    \item Develop a user-friendly interface with interactive elements such as graphs
    \item Design the detection algorithm to be language-agnostic and develop language-specific adapters
    \item Provide CLIs and Application Programming Interface (API) endpoints for seamless integration with development tools
    \item Use a modular architecture for scalability and regular updates for maintainability
    \item Keep data processing local to the user’s environment and implement robust encryption for cloud processing
    \item Keep updated with the latest software development trends and encourage community engagement
\end{enumerate}

\section{Discussion}

It is more efficient to make project changes in one place than to create updates across multiple IDE plugins or systems. Therefore, we propose encapsulating these functions within packages. Given the potential for local web support, we consider NPM packages an effective means of achieving this objective. We recognize that the seamless integration of NPM packages into various IDEs, including IntelliJ, is particularly beneficial. However, this approach has several challenges because it requires merging two distinct runtime environments.

Incorporating a feature that visualizes data clumps directly within an IDE could significantly assist users in several ways. For example, during the development process, users could instantly ascertain whether their current work has any dependencies or relationships with files located elsewhere in the project. For existing projects with numerous data clumps, users could employ an interactive graph to navigate to files of interest. Additionally, integrating an interactive graph into an IDE could help users select specific methods, classes, or fields for automatic refactoring, increasing users’ autonomy in the refactoring phase. In terms of accessibility and ease of use, a standalone web application could be accessed from any device with a web browser, making it more accessible to users who do not have a specific IDE installed. This approach could be particularly beneficial for teams using different development environments or individuals who need to review the data clusters without accessing the full development setup. We believe that integration into IntelliJ would be helpful.

Palomba et al. \cite{HIST} created a framework for examining the evolution of code smells over time. It would be beneficial to use this information when refactoring data clumps. Segregating functionalities into different modules, such as those dedicated to visualization, helps users incorporate the functionalities into continuous integration pipelines or other web-based tools, such as Qoodana. Srategic integration into automated workflows could mitigate the risks commonly associated with software development \cite{duvall2007continuous}.

Another solution is to offer detection and refactoring as a cloud service. However, we have chosen not to pursue this option. Keeping the analysis within the project’s environment ensures that sensitive data does not leave the organizational boundary, reducing the risk of security breaches. This approach is particularly suitable for organizations governed by strict data protection regulations, as the detection and refactoring processes remain local. Local processing also eliminates dependence on external services, ensuring that detection and refactoring are not impeded by network issues or service downtime. For example, a company working on confidential software projects risks exposing sensitive information when using an external service for data clump detection and refactoring. In contrast, our approach confines the entire process to the company’s internal development environment, ensuring that sensitive data, such as proprietary algorithms or customer information embedded in the code, remains within the secure confines of the company’s network.

\section{Conclusion}

Not integrating data clump detection and refactoring workflows exclusively into IDEs has advantages and disadvantages. We believe that centralized control of the detection algorithm is a key advantage, especially for future applications in different environments and programming languages because it simplifies the tasks of maintenance and further improvement. The simplicity of integrating NPM packages into an IDE plugin is a feature we consider desirable.

Initially, we focused on creating a plugin dedicated to IDE integration, ensuring users received instant feedback during the development process. However, in our updated approach, we highlight the potential of a tool that automatically identifies and refactors data clumps in public repositories, functioning similarly to an external contributor. This direction maintains our original goal while offering the added advantage of automated checks, improving code quality and project management.



\begin{thebibliography}{99}

\bibitem{improveSoftwareQualityThroughAutomation}
Mohammad, Sikender Mohsienuddin, ``Improve Software Quality Through Practicing DevOps Automation,'' \emph{International Journal of Creative Research Thoughts (IJCRT)}, vol. 6, no. 1, pp. 251--256, March 2018. [Online]. Available: \url{http://www.ijcrt.org/papers/IJCRT1133482.pdf}.

\bibitem{howWeRefactor}
Murphy-Hill, Emerson, Parnin, Chris, and Black, Andrew P., ``How We Refactor, and How We Know It,'' \emph{IEEE Transactions on Software Engineering}, vol. 38, no. 1, pp. 5-18, 2012. DOI: \url{10.1109/TSE.2011.41}.

\bibitem{automateRefactorings}
Golubev, Yaroslav, Kurbatova, Zarina, AlOmar, Eman Abdullah, Bryksin, Timofey, and Mkaouer, Mohamed Wiem, ``One Thousand and One Stories: A Large-Scale Survey of Software Refactoring,'' in \emph{Proceedings of the 29th ACM Joint Meeting on European Software Engineering Conference and Symposium on the Foundations of Software Engineering}, ESEC/FSE 2021, Athens, Greece, 2021, pp. 1303--1313, New York, NY, USA: Association for Computing Machinery. DOI: \url{10.1145/3468264.3473924}.

\bibitem{intellijpluginframework}
Kurbatova, Z., Golubev, Y., Kovalenko, V., and Bryksin, T., ``The IntelliJ Platform: A Framework for Building Plugins and Mining Software Data,'' in \emph{2021 36th IEEE/ACM International Conference on Automated Software Engineering Workshops (ASEW)}, Los Alamitos, CA, USA: IEEE Computer Society, Nov 2021, pp. 14--17. DOI: \url{10.1109/ASEW52652.2021.00016}.

\bibitem{duvall2007continuous}
Duvall, Paul, Matyas, Steve, and Glover, Andrew, \emph{Continuous Integration: Improving Software Quality and Reducing Risk}, Addison-Wesley Professional, 2007, ISBN: 9780321336385.

\bibitem{scriptingParametricRefactorings}
Kim, Jongwook, Batory, Don, and Dig, Danny, ``Scripting parametric refactorings in Java to retrofit design patterns,'' 2015. DOI: \url{10.1109/ICSM.2015.7332467}.

\bibitem{influenceOfHumanFactorsInSoftwareQuality}
Fern{\'a}ndez-Sanz, Luis and Misra, Sanjay, ``Influence of Human Factors in Software Quality and Productivity,'' in \emph{Computational Science and Its Applications - ICCSA 2011}, Springer Berlin Heidelberg, Berlin, Heidelberg, 2011, pp. 257--269. DOI: \url{10.1007/978-3-642-21934-4_22}.

\bibitem{ImprovedCodeSmellDefinition}
Zhang, Min, Baddoo, Nathan, Wernick, Paul, and Hall, Tracy, ``Improving the Precision of Fowler's Definitions of Bad Smells,'' \emph{2008 32nd Annual IEEE Software Engineering Workshop}, Nov 2008, pp. 161--166. DOI: \url{10.1109/SEW.2008.26}.

\bibitem{HIST}
Palomba, Fabio, Bavota, Gabriele, Penta, Massimiliano Di, Oliveto, Rocco, Poshyvanyk, Denys, and De Lucia, Andrea, ``Mining Version Histories for Detecting Code Smells,'' \emph{IEEE Transactions on Software Engineering}, vol. 41, no. 5, pp. 462-489, 2015. DOI: \url{10.1109/TSE.2014.2372760}.

\bibitem{PMD}
PMD, ``PMD - An extensible cross-language static code analyzer,'' 2023. [Online]. Available: \url{https://pmd.github.io/}. Accessed: Oct. 08, 2023.

\bibitem{baumgartner2023live}
Baumgartner, Nils, Adleh, Firas, and Pulverm{\"u}ller, Elke, ``Live Code Smell Detection of Data Clumps in an Integrated Development Environment,'' in \emph{Proceedings of the 18th International Conference on Evaluation of Novel Approaches to Software Engineering}, Prague, Czech Republic, April 2023, pp. 64--76, SciTePress. DOI: \url{10.5220/0011727500003464}.

\bibitem{CostOfMaintenance}
Brown, William H., Malveau, Raphael C., McCormick, Hays W., and Mowbray, Thomas J., \emph{AntiPatterns: Refactoring Software, Architectures, and Projects in Crisis}, 1st ed., John Wiley \& Sons, Inc., USA, 1998, ISBN: 0471197130.

\bibitem{Fowler1999}
Becker, Paul, Fowler, Martin, Beck, Kent, Brant, John, Opdyke, William, and Roberts, Don, \emph{Refactoring - Improving the Design of Existing Code}, Addison-Wesley Professional, Boston, 1999, ISBN: 978-0-201-48567-7.

\bibitem{refactoringBot}
Wyrich, Marvin and Bogner, Justus, ``Towards an Autonomous Bot for Automatic Source Code Refactoring,'' in \emph{2019 IEEE/ACM 1st International Workshop on Bots in Software Engineering (BotSE)}, 2019, pp. 24--28. DOI: \url{10.1109/BotSE.2019.00015}.

\end{thebibliography}


\end{document}